\documentclass[twocolumn,twoside,slac_two]{revtex4}
\usepackage{graphicx}
\usepackage{fancyhdr}
\renewcommand{\headrulewidth}{0pt}
\renewcommand{\footrulewidth}{0pt}

\def\beq{\begin{equation}}
\def\eeq{\end{equation}}

\begin{document}

\title{Force-Free Magnetosphere of an Accreting Kerr Black Hole}

%

\author{Dmitri A. Uzdensky}
\affiliation{Princeton University, Princeton, NJ 08540, USA}

\begin{abstract}
I consider a stationary axisymmetric force-free degenerate
magnetosphere of a rotating Kerr black hole surrounded by a
thin Keplerian infinitely-conducting accretion disk. I focus
on the closed-field geometry with a direct magnetic
coupling between the disk and the event horizon. I first
present a simple physical argument that shows how the black
hole's rotation limits the radial extent of the force-free link.
I then confirm this result by solving numerically the
general-relativistic force-free Grad--Shafranov equation
in the magnetosphere, using the regularity condition at
the inner light cylinder to determine the poloidal current.
I indeed find that force-free solutions exist only when
the magnetic link between the hole and the disk has a limited
extent on the disk surface. I chart out the maximum allowable
size of this magnetically-connected part of the disk as a function
of the black hole spin. I also compute the angular momentum and
energy transfer between the hole and the disk that takes place
via the direct magnetic link. I find that both of these
quantities grow rapidly and that their deposition becomes
highly concentrated near the inner edge of the disk as the
black hole spin is increased.
\end{abstract}

\maketitle

\thispagestyle{fancy}



\section{Introduction}
\label{sec-intro}

This paper is a short version of the full article~\cite{Uzdensky-2005}.

Magnetic fields play a crucial role in both the dynamics and
the energetics of accretion flows onto black holes, which makes
it important to understand the global structure of an accreting 
black hole magnetosphere.

Conceptually, one can think of two basic types of magnetic-field geometry.
The first type is the open-field configuration responsible 
for the Blandford--Znajek mechanism for powering AGN jets~\cite{BZ77}. 
In this configuration there is no direct magnetic link 
between the hole and the disk: all the field lines are open 
and extend from the black hole and from the disk to infinity.
The open magnetic field extracts the rotational energy
and angular momentum from a spinning black hole and transports 
them outward to power a jet. A similar process also works for the 
field lines connected to the disk.

\begin{figure}[t]
\includegraphics[width=2.5in]{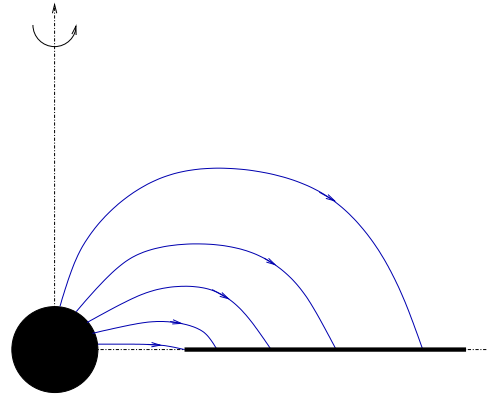}
\caption{Schematic drawing of a {\it fully-closed} 
black hole -- disk magnetosphere. Energy and angular 
momentum are exchanged between the hole and the disk
through a direct magnetic link.}
\label{fig-geometry-closed}
\end{figure}

Recently, there's been a rising interest in a different, namely, 
closed-field configuration (Figure~\ref{fig-geometry-closed}), 
characterized by a direct magnetic coupling between the hole and the 
disk~\cite{Blandford-1999,Gruzinov-1999,Li-2000,Li-2002a,Li-2002b,Li-2004,Wang-2002,Wang-2003,Wang-2004,vanPutten-Levinson-2003,Uzdensky-2004,Uzdensky-2005}.
Here, energy and angular momentum are directly exchanged between 
the hole and the disk via the magnetic field. This exchange governs 
the spin evolution of the black hole~\cite{Wang-2002,Wang-2003,Wang-2004} 
and also has a strong effect on the energy dissipation profile in the disk, 
which leads to important observational consequences~\cite{Li-2002b,Li-2004}. 
In order to understand any of these processes, however, one needs 
to have a good picture of the  structure of the magnetic field. 
Whereas there have been some numerical studies of open-field 
force-free configurations~\cite{Macdonald-1984,Fendt-1997,Komissarov-2001,Komissarov-2002,Komissarov-2004}, 
the structure of the  closed-field magnetosphere 
have remained essentially unexplored, with the exception of my recent 
work on the Schwarzschild case~\cite{Uzdensky-2004}. 
This provides the motivation for my present study~\cite{Uzdensky-2005}.

\section{The Model}
\label{sec-model}

The main goal of this work is to determine the structure  
of the coupled disk--black hole magnetosphere under the 
following assumptions:
1) everything is axially-symmetric and stationary;
2) the magnetosphere is force-free and infinitely-conducting;
3) the disk is infinitely conducting, thin, and Keplerian.

The structure of such a magnetosphere is governed by 
the generally-relativistic force-free Grad--Shafranov 
(e.g., \cite{MT82,Beskin-1997}) for the poloidal flux 
function~$\Psi$ (in Boyer-Lindquist coordinates):
\begin{eqnarray}
\nabla \cdot \biggl( {\alpha\over{\varpi^2}}\, 
\bigl[ 1-{{\Delta\Omega^2 \varpi^2}\over{\alpha^2}} \bigr] 
\nabla\Psi \biggr) &+& \nonumber \\
{{\Delta\Omega}\over\alpha}\, {d\Omega_F\over{d\Psi}}\, (\nabla\Psi)^2 +
{1\over{\alpha\varpi^2}}\, II'(\Psi) &=& 0 \, .
\label{eq-GS}
\end{eqnarray}
Here, $\alpha$ is the lapse function, $\varpi$ is the cylindrical radius,
$\Omega_F(\Psi)$ is the field-line angular velocity, 
$\Delta\Omega = \Omega_F-\omega$, where $\omega$ is 
the frame-dragging frequency, and $I=I(\Psi)$ is the poloidal current.

This equation is very non-trivial, 
not only because it's a non-linear partial differential 
equation, but also because it involves two, in general 
nonlinear, functions of $\Psi$, namely $\Omega_F(\Psi)$ 
and $I(\Psi)$, that somehow need to be specified. 

In our particular problem, the first of these functions, 
$\Omega_F(\Psi)$, is determined fairly straight-forwardly 
from the boundary conditions. Indeed, all the field lines
threading the hole are also attached to an infinitely 
conducting Keplerian disk. Therefore, in steady state, 
the angular velocity of any given field line $\Psi$ 
is equal to the Keplerian angular velocity of this 
line's footpoint $r_0(\Psi)$ on the surface of the disk:
$\Omega_F(\Psi)=\Omega_K[r_0(\Psi)]$. 
Thus, if a Dirichlet boundary condition for equation~(\ref{eq-GS}) 
on the disk surface is specified as $\Psi=\Psi_d(r)$, 
one can invert $\Psi_d(r)$ and obtain an explicit 
expression for $\Omega_F(\Psi)$.

The poloidal current $I(\Psi)$, on the other hand, cannot be 
determined as easily, since there is no surface on which one 
can just explicitly prescribe $I(\Psi)$.
Instead, one has to use an indirect procedure that
makes use of the regularity condition at the inner 
light cylinder. Indeed, as has first been noticed by 
Znajek~\cite{Znajek-1977} and by Blandford \& Znajek~\cite{BZ77},
any rotating field line going into the hole has to pass
through an inner Light Cylinder (LC), just as any rotating 
field line going out to infinity has to pass through an outer~LC. 
The light cylinder is defined as the surface where
the rotational velocity of field lines, as measured by
a local Zero-Angular-Momentum Observer (ZAMO), is equal 
to the speed of light and where  $E=B_{\rm pol}$  in the 
ZAMO frame:
\beq
\alpha=\alpha_{\rm LC} = |\Delta\Omega| \varpi\, , 
\label{eq-LC}
\eeq
(in our problem, this inner LC is not at all a cylinder but 
rather an almost spherical surface between the event horizon
and the boundary of the ergosphere). As one can easily see,
the LC is a regular singular surface of equation~(\ref{eq-GS}). 
This means that, in general, this equation admits solutions 
that are not continuous or continuously differentiable at 
$r=r_{\rm LC}(\theta)$. That is, the regularity of the solution 
at the LC does not follow from the equation automatically, but 
has to be imposed as an additional physical condition. 
And it is precisely this condition that one can use to 
uniquely fix the function $I(\Psi)$!
Actually doing this numerically requires an iterative approach 
in which $I(\Psi)$ is evolved together with $\Psi(r,\theta)$
until a regular solution of equation~(\ref{eq-GS}) is obtained.
This is very similar in spirit to the procedure used by 
Contopoulos~et~al.~\cite{CKF99} in their study of a force-free 
magnetosphere of an axisymmetric pulsar 
(see also~\cite{Uzdensky-2003,Uzdensky-2004}). 
One essential complication in the black hole case is that 
the field lines rotate differentially and hence the actual 
location and shape of the LC are not known a priori. 
Therefore,  one cannot set up the computational grid 
in a manner convenient for treating the LC regularity
condition; in general, the LC sweeps through the grid 
until the final solution is achieved and so one has to 
actually evolve $r_{\rm LC}(\theta)$ dynamically during 
the iteration procedure.

Now let me discuss the role of the black hole event horizon, 
$\alpha=0$. The event horizon is also a regular singular 
surface of the Grad-Shafranov equation~(\ref{eq-GS}). 
This is a very important observation; in particular, 
it means that one cannot, and indeed need not, provide 
any additional independent boundary condition here. 
Instead, one should impose the condition of regularity 
of the solution at the horizon. This condition can be 
actually obtained by simply setting $\alpha=0$ in 
equation~(\ref{eq-GS}), with the following result,
commonly known as Znajek's event-horizon boundary 
condition~\cite{Znajek-1977} (see also~\cite{MT82}):
\beq
I[\Psi_0(\theta)]=
{{2Mr_H\sin\theta}\over{\rho^2}}\, \Delta\Omega\, {d\Psi_0\over{d\theta}}\, ,
\qquad r=r_H \, ,
\label{eq-EH-bc}
\eeq
where $\Psi_0(\theta)\equiv \Psi(r=r_H,\theta) $ is the horizon 
magnetic flux distribution. 
In fact, this condition is what the fast magnetosonic critical
condition becomes in the limit when the plasma density is taken 
to zero and the fast magnetosonic surface approaches the horizon
\cite{Beskin-1997,BK-2000,Komissarov-2004}.
The significance of condition~(\ref{eq-EH-bc}) is that it can be 
used to determine the magnetic flux distribution $\Psi_0(\theta)$ 
at the horizon provided that the functions $I(\Psi)$ and 
$\Omega(\Psi)$ are already known.


\section{Main Effect: Disruption of the Hole--Disk Coupling by 
The Black Hole's Rotation}
\label{sec-idea}

Another reason the event horizon regularity condition is very important 
is that it enables us to arrive at a very important effect limiting the 
maximal possible spatial extent of a direct force-free magnetic link 
between a rotating black hole and a conducting Keplerian disk. 
Here I present the basic physical reason for this effect 
(confirmed by the numerical simulations presented in the next section).

First, let us suppose that a force-free configuration of Figure~\ref
{fig-geometry-closed}, where all the field lines attached to 
the disk at arbitrarily large radii thread the event horizon, 
does indeed exist.
Then let us consider the polar region of the black hole, $r=r_H$, 
$\theta\rightarrow 0$, and suppose that near the rotation axis the flux
$\Psi_0(\theta)$ behaves as a power law: $\Psi_0\sim\theta^\gamma$ 
(the most natural choice being $\Psi_0\sim\theta^2$). 
Next, note that the field lines threading the horizon in this polar 
region connect to the disk at some very large radii, $r_0(\Psi)\gg r_H$. 
For sufficiently small~$\Psi$, and hence sufficiently large $r_0(\Psi)$, 
the Keplerian angular velocity is small, and therefore $\Omega_F(\Psi)$ 
is much smaller than the black hole rotation rate $\Omega_H = a/2r_H$, 
so that $\Delta\Omega\approx -\Omega_H={\rm const}$.
Then, from the event horizon regularity condition~(\ref{eq-EH-bc})
it follows that $I(\Psi) \sim -\Omega_H \Psi \sim -a\Psi$, as 
$\Psi\rightarrow 0$, and so
\beq
II'(\Psi\rightarrow 0) \sim a^2 \Psi \, .
\label{eq-II'-axis}
\eeq

Now, if we look at the force-free balance on the same field lines
but far away from the black hole, at distances comparable with the
footpoint radius, we can estimate the linear term on the left-hand-side 
of the Grad--Shafranov simply as $\Psi/r^2$. We see that both the 
left-hand side and the right-hand side given by equation~(\ref{eq-GS}) 
scale linearly with~$\Psi$ but the left-hand side has an additional 
factor $\sim r^{-2}$. Thus, at sufficiently large distances this term 
becomes negligible when compared with the $II'(\Psi)$-term given by 
(\ref{eq-II'-axis}). 
In other words, the toroidal field produced in the polar region
of the horizon by the black hole dragging the field lines is too
large to be confined by the poloidal field tension at large enough
distances. 
In fact, this argument suggests that the maximal radial extent $R_{\rm max}$ 
of the region on the disk connected to the polar region of a Kerr black 
hole should scale as $R_{\rm max}\sim r_H/a$ in the limit $a\rightarrow 0$.
In the Schwarzschild limit $a\rightarrow 0$, this 
maximal distance goes to infinity and hence a fully-closed force-free 
configuration can exist at arbitrarily large distances, in agreement 
with my previous conclusions~\cite{Uzdensky-2004}. 

To sum up, even though the field lines can, to a certain degree, 
slip through the horizon because the latter is effectively resistive
in the language of the Membrane Paradigm~\cite{MP86}, in some situations 
the horizon is just not resistive enough! 
Indeed, the field lines are "dragged" by the rotating black hole to such 
a degree that, in order for them to slip through the horizon steadily, 
they must have a certain, rather large toroidal field component. When, 
for fixed disk boundary conditions, the black-hole spin parameter~$a$ 
is increased beyond a certain limit~$a_{\rm max}$, this toroidal field 
becomes so large that the poloidal field tension is no longer able to 
contain its pressure at large distances.


\section{Numerical Results}
\label{sec-numerical}

In order to verify the proposition put forward in the preceding
section, I have performed a series of numerical calculations and
obtained solutions of the force-free Grad--Shafranov equation for 
various values of two parameters: the black-hole spin~$a$ and the 
radial extent $R_s$ of the magnetic link. In this section I describe 
the computational set-up of the problem and present the main numerical 
results.

The simplest axisymmetric closed-field configuration one could
consider is that shown in Figure~\ref{fig-geometry-closed}, where
all magnetic field lines connect the disk and the hole. Furthermore, 
the entire event horizon and the entire disk surface participate in 
this magnetic linkage; in particular, the field lines threading the 
horizon very close to the axis $\theta=0$ are anchored at some very 
large radial distances in the disk. 
However, as follows from the arguments presented in \S~\ref{sec-idea},
a steady-state force-free configuration of this type can only exist 
in the case of a Schwarzschild black hole. This is in accord with my 
simulations, as I was not able to obtain a convergent solution even 
for a Kerr black hole with the spin parameter as small as $a=0.05$.

From the discussion in \S~\ref{sec-idea}, we expect that, for a given 
value of~$a$, the magnetic link between the polar region of the black 
hole and the disk cannot extend to distances on the disk larger than 
a certain $R_{\rm max}(a)$.  Although the actual dependence $R_{\rm max}(a)$ 
should depend on the details of the problem, there are some general properties 
that it should have; in particular, one expects $R_{\rm max}$ to be a 
monotonically decreasing function of~$a$ and to be inversely proportional 
to~$a$ in the limit $a\rightarrow 0$.

\begin{figure}
\includegraphics[width=2.5in]{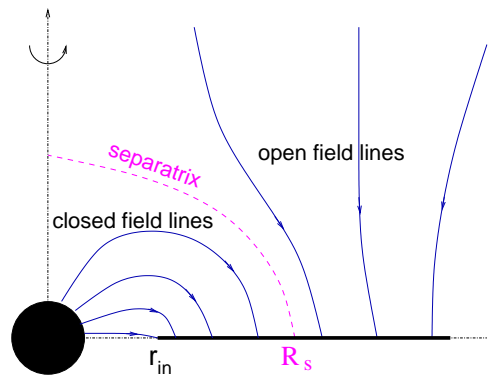}
\caption{Schematic drawing of a black hole -- disk magnetosphere
with a {\it radially-limited magnetic connection}. Here, only the
inner part of the disk is coupled magnetically to the hole (closed
field region), whereas the field lines attached to the outer part 
of the disk are open and extend to infinity.}
\label{fig-geometry-kerr}
\end{figure}

In order to test this picture and to determine the actual functional shape 
of $R_{\rm max}(a)$, I modified the basic geometry of the configuration by 
allowing for two distinct regions: the region of closed field lines 
connecting the black hole to the inner part of the disk $r<R_s$, and 
the region of open field lines extending from the outer part of the disk 
all the way to infinity (see Fig.~\ref{fig-geometry-kerr}). 
Thus, there is a critical field line, $\Psi_s\equiv \Psi_d(R_s)
<\Psi_{\rm tot}$, that separates open field lines ($\Psi<\Psi_s$)
from the closed field lines ($\Psi_s<\Psi<\Psi_{\rm tot}$) connecting 
to the black hole (I count the poloidal flux on the disk surface from 
the radial infinity inward).
Correspondingly, the poloidal flux on the black hole surface varies 
from $\Psi=\Psi_s$ at the pole $\theta=0$ to $\Psi=\Psi_{\rm tot}$ 
at the equator $\theta=\pi/2$.

Let me now describe the computational domain and the boundary conditions.
First, because of the assumed axial symmetry and the symmetry with respect 
to the equatorial plane, I performed the computations only in one quadrant: 
$\theta\in[0,\pi/2]$ and $r\in[r_H,\infty]$. Thus, there are four boundaries:
the axis $\theta=0$, the infinity $r=\infty$, the equator $\theta=\pi/2$, 
and the horizon $r=r_H$. Of these, the axis and the equator require boundary 
conditions for~$\Psi$, whereas the horizon and the infinity are actually 
regular singular surfaces and so one can only impose regularity conditions 
on them.

The boundary condition on the axis is simply
\beq
\Psi(r,\theta=0) = \Psi_s = {\rm const} \, .
\label{eq-bc-axis}
\eeq

The equatorial boundary, $\theta=\pi/2$, actually consists of two parts: 
the infinitesimally thin disk and the plunging region between the inner 
edge of disk, $r_{\rm in}=r_{\rm ISCO}(a)$, and the black hole.
On the disk surface, $r>r_{\rm in}$, I impose a Dirichlet-type
boundary condition:
\beq
\Psi(r>r_{\rm in},\theta=\pi/2) \equiv \Psi_d(r)= 
\Psi_{\rm tot}\, {{r_{\rm in}}\over r} \, .
\label{eq-bc-disk}
\eeq

In the plunging region, $r_H\leq r\leq r_{\rm in}$, I set
\beq
\Psi(r_H\leq r\leq r_{\rm in},\theta=\pi/2)=\Psi_{\rm tot} \equiv
\Psi_d(r_{\rm in}) = {\rm const} \, .
\label{eq-bc-plunging}
\eeq
This choice appears to be physically appropriate for an accreting 
disk, because the matter in this region falls rapidly onto the black 
hole and stretches the magnetic loops in the radial direction, 
greatly diluting the vertical field component. The horizontal 
magnetic field reverses across the plunging region, which is 
therefore represented by an infinitesimally thin non-force-free
current sheet. 

Next, as I have discussed in~\S~\ref{sec-model}, the event horizon 
is a regular singular surface of the Grad--Shafranov equation and 
so I impose the regularity condition~(\ref{eq-EH-bc}) there. 
This condition is an ordinary differential equation that determines 
the function~$\Psi_0(\theta)$ provided that both $\Omega_F(\Psi)$ 
and $I(\Psi)$ are given. Thus, from the procedural perspective, 
it can be viewed as a Dirichlet boundary condition on the horizon. 

Similarly, the spatial infinity $r=\infty$ is also a regular singular 
surface of the Grad--Shafranov equation and thus can also be described 
by a regularity condition. In this sense, the horizon and the infinity 
are equivalent~\cite{Punsly-Coroniti-1990}. 
In the particular problem set-up considered here, I 
set $\Omega_F(\Psi)=0$ and $I(\Psi)=0$ 
on the open field lines going from the disk
to infinity (see below). Then the Grad--Shafranov 
equation at large distances is greatly simplified
and the infinity regularity condition yields
\beq
\Psi(r=\infty,\theta)=\Psi_s \cos\theta \, .
\label{eq-bc-infty}
\eeq

In addition to boundary conditions, one has to specify the 
angular velocity $\Omega_F(\Psi)$ of the magnetic field lines,
which in principle should be equal to the Keplerian angular 
velocity in the disk:
$\Omega_F(\Psi)=\Omega_K[r_0(\Psi)]=\sqrt{M}/(r_0^{3/2}+a\sqrt{M})$.
However, for simplicity I set $\Omega_F(\Psi)\equiv 0$ for
the open field lines extending to infinity, $\Psi<\Psi_s$.
This way I don't have to deal with the outer light cylinder 
that these lines would have to cross; correspondingly, I also
set $I(\Psi<\Psi_s)\equiv 0$. 
In other words, I assume that the outer, open-field portion 
of the disk is not rotating and that the open-field part of 
the magnetosphere attached to it is potential.
Next, in order to avoid the discontinuities in $\Omega_F(\Psi)$ 
and $I(\Psi)$ at $\Psi=\Psi_s$, I modify the disk rotation law 
slightly just inside of $R_s$ by taking $\Omega_F$ smoothly to
zero over a small vicinity of the separatrix. 
These modifications enabled me to focus on examining the 
effect of the black hole rotation on the maximal allowable 
radial extent $R_s$ of the force-free magnetic coupling,
while avoiding the numerical difficulties resulting from 
the discontinuous behavior of $I(\Psi)$, etc. I believe 
that these modifications do not lead to any qualitative 
change in the conclusions, especially in the case of small~$a$.

\begin{figure}[t]
\includegraphics[width=2.5in]{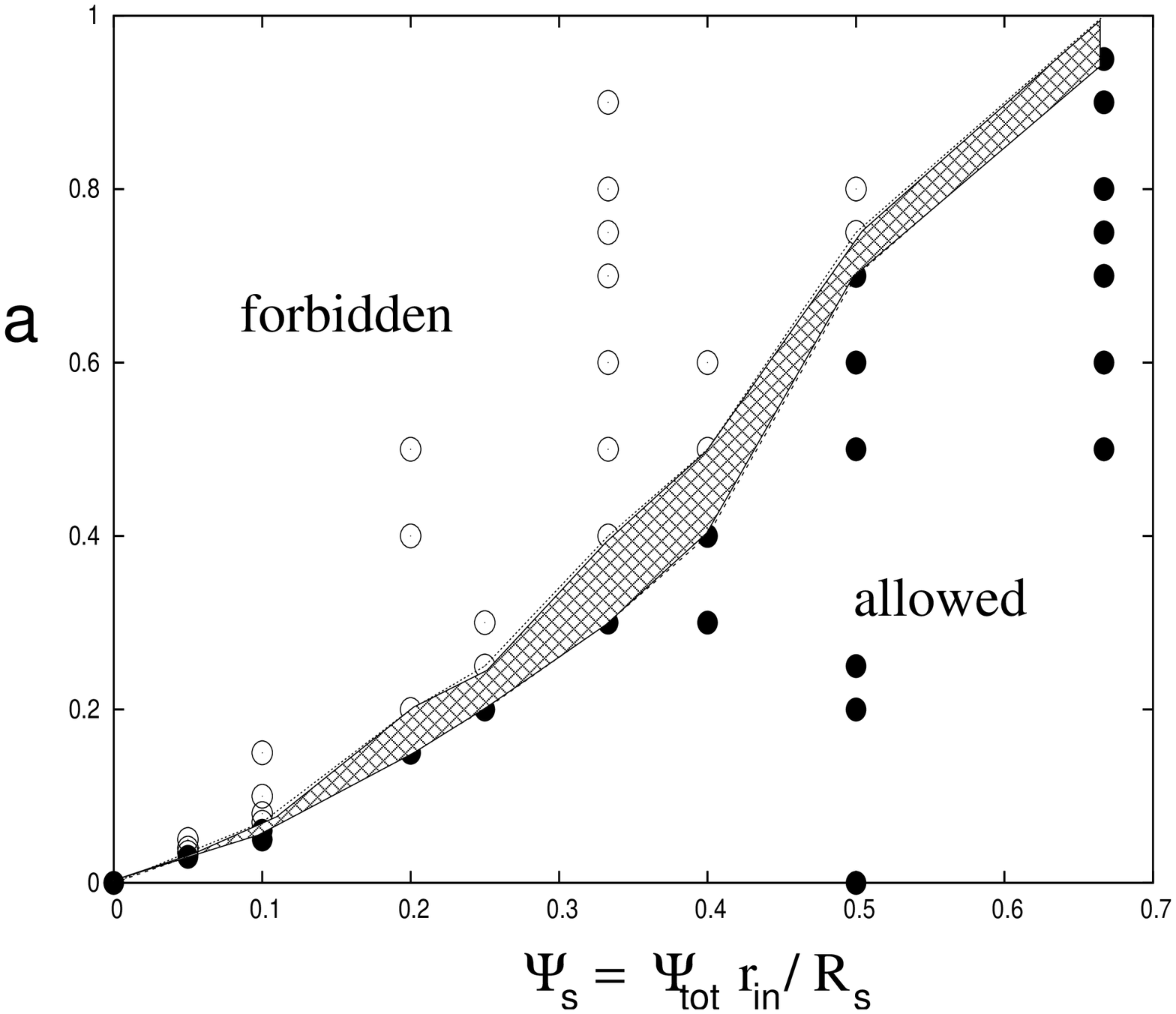}
\caption{The 2-D parameter space $(a,\Psi_s)$. 
The shaded band running diagonally across the plot represents
the function $a_{\rm max}(R_s)$, the maximal value of $a$ for 
which a force-free magnetic link can extend up to a given radial 
distance $R_s$ on the disk.}
\label{fig-a_of_Psi_s}
\end{figure}

The single most important result of my calculations is presented 
in Figure~\ref{fig-a_of_Psi_s}. This figure shows where in the 
two-dimensional $(a,\Psi_s)$ parameter space force-free solutions 
exist. Filled circles on this plot represent the runs in which 
convergence was achieved (allowed region), whereas open circles 
correspond to the runs that failed to converge to a suitable solution 
(forbidden region). The boundary $a_{\rm max}(\Psi_s)$ between the 
allowed and forbidden regions is located somewhere inside the narrow 
hatched band that runs from the lower left to the upper right of the 
Figure. As one can see, $a_{\rm max}(\Psi_s)$ is a monotonically 
increasing function and it indeed scales linearly with $\Psi_s$ 
and hence inversely with $R_s\sim 1/\Psi_s$, in the limit $\Psi_s 
\rightarrow 0$, in full agreement with the arguments of \S~\ref{sec-idea}.

\begin{figure}[t]
\includegraphics[width=3in]{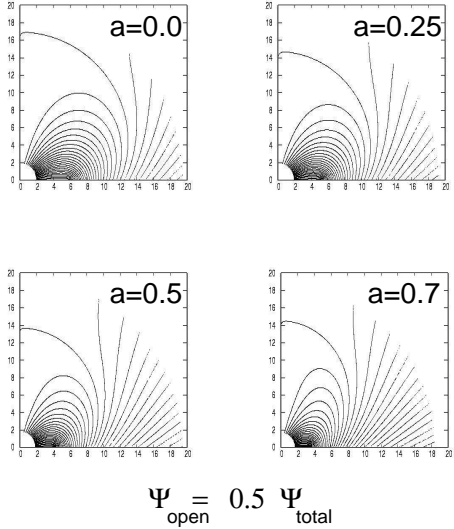}
\caption{Contour plots of the magnetic flux function $\Psi(r,\theta)$
for $a=0.0$, 0.25, 0.5, and~0.7; the total open flux is 
$\Psi_s=0.5\Psi_{\rm tot}$ in all cases, corresponding to 
$R_s=2r_{\rm in}$.}
\label{fig-contour}
\end{figure}

In order to study the effect that the black hole spin has on the solutions, 
I concentrate on several values of~$a$ for a fixed value of~$\Psi_s$. 
In particular, I choose $\Psi_s=0.5$, which corresponds to $R_s=
2r_{\rm in}(a)$, and consider four values of~$a$: $a=0$, 0.25, 0.5, and~0.7. 
Figure~\ref{fig-contour} shows the contour plots of the poloidal magnetic 
flux for these four cases. One can see that although the flux surfaces 
inflate somewhat with increased~$a$, this expansion is not very dramatic, 
even in the case $a=0.7$, which is very close to the critical value 
$a_{\rm max}(\Psi_s=0.5)$ at which a sudden loss of equilibrium occurs.

\begin{figure}[t] 
\hskip -83pt
\includegraphics[width=2.5in]{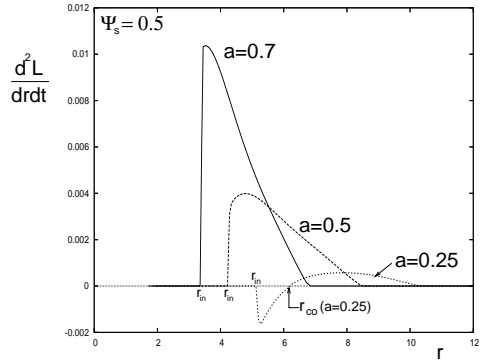}
\caption{Radial distribution of the magnetic torque 
per unit radius~$r$ on the disk surface for $a=0.25$, 
0.5, and~0.7.}
\label{fig-L-of-r}
\end{figure}

\begin{figure}
\hskip -83pt
\includegraphics[width=2.5in]{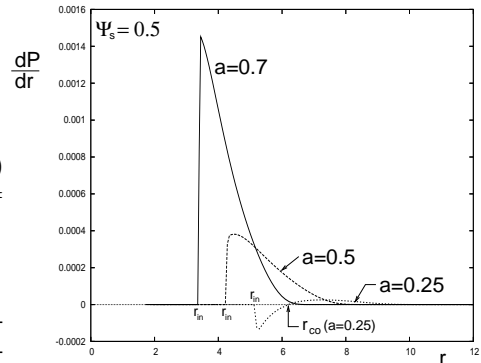}
\caption{Radial distribution $dP/dr$ of the red-shifted power 
per unit radius~$r$ on the disk surface for $a=0.25$, 0.5, and~0.7.}
\label{fig-P-of-r}
\end{figure}

After the solutions have been obtained, one can compute
the rate of exchange of energy and angular momentum between 
the black hole and the disk. According to \cite{MT82}, 
the amount of angular momentum $\Delta L$ transported out 
in a unit of global time $t$ through a flux tube between 
$\Psi$ and $\Psi+\Delta\Psi$ is $-I(\Psi)\Delta\Psi/2$
and the red-shifted power is $-I(\Psi)\Omega_F(\Psi)\Delta\Psi/2$.
Then, since in our problem we have an explicit mapping~(\ref{eq-bc-disk}) 
between $\Psi$ and~$r$, we can obtain the radial distributions of 
the deposition of angular momentum and red-shifted energy on the disk. 
Figures~\ref{fig-L-of-r} and~\ref{fig-P-of-r} show these distributions 
for our selected cases $a=0.25$, 0.5, and~0.7 for fixed $\Psi_s=0.5$. 
We see that in the case $a=0.25$ there is a corotation point $r_{\rm co}$
on the disk such that $\Omega_{\rm disk}>\Omega_H$ inside $r_{\rm co}$ 
and $\Omega_{\rm disk}<\Omega_H$ outside $r_{\rm co}$. Correspondingly,
both angular momentum and red-shifted energy flow from the inner
($r<r_{\rm co}$) part of the disk to the black hole and from the 
hole to the outer ($r>r_{\rm co}$) part of the disk. At larger values 
of~$a$, however, the Keplerian angular velocity at $r=r_{\rm in}$ is 
smaller than the black hole's rotation rate and there is no corotation 
point; correspondingly, both angular momentum and red-shifted energy 
flow from the hole to the disk. Also, as can be seen in Figures~\ref
{fig-L-of-r} and~\ref{fig-P-of-r}, the deposition of these quantities 
becomes strongly concentrated near the disk's edge, especially at higher
values of~$a$.

A more detailed description of the numerical model and 
of the results is given in~\cite{Uzdensky-2005}.


\section{Conclusions}
\label{sec-conclusions}

I investigated the structure of a force-free magnetosphere linking
a rotating Kerr black hole to its accretion disk. I assumed that 
the magnetosphere is stationary, axisymmetric, and infinitely-conducting 
and that the disk is thin, ideally conducting, and Keplerian. 
My main goal was to determine under what conditions a force-free 
magnetic field can connect the hole directly to the disk and how 
the black hole rotation limits the radial extent of such a link.

I first introduced a simple physical argument demonstrating that 
magnetic field lines connecting the polar region of a spinning 
black hole to arbitrarily remote regions of the disk cannot be 
in a force-free equilibrium.
The basic reason for this is that these field lines would have 
to have a substantial toroidal magnetic field component, which, 
in the language of the Membrane Paradigm~\cite{MP86}, is needed 
for the field lines to slip resistively across the stretched 
event horizon. 
In a force-free magnetosphere, toroidal flux spreads along the lines 
to keep the poloidal current constant along the field. 
Then one finds that the outward pressure of this toroidal field 
is too large to be confined by the poloidal field tension at large 
distances from the hole. 
These means that these field lines cannot be in a force-free equilibrium. 
Furthermore, one can generalize this argument to the case of closed 
magnetospheres of finite size and derive a conjecture that the maximal 
radial extent $R_{\rm max}$ of the magnetically-coupled region on the disk 
surface should scale inversely with the black hole spin parameter~$a$ in 
the limit $a\rightarrow 0$.

I have then obtained numerical solutions of the general-relativistic 
force-free Grad--Shafranov equation that governs the system's behavior.
To do this, I had to devise an iterative procedure that used the inner 
light cylinder regularity condition to determine the poloidal 
current~$I(\Psi)$ and the shape of the light cylinder simultaneously 
with calculating the solution~$\Psi(r,\theta)$. 
I performed a series of computations corresponding to various values of 
two parameters: the black hole spin $a$ and the magnetic link's radial 
extent $R_s$ on the disk surface.
I was able to chart out the boundary $a_{\rm max}(\Psi_s)$ between the 
allowed and the forbidden domains in the two-parameter space $(a,\Psi_s)$. 
I found that this is a monotonically rising curve with the asymptotic 
behavior $a_{\rm max}\propto \Psi_s$ as $\Psi_s\rightarrow 0$, 
in line with the predictions of \S~\ref{sec-idea}.
I also computed the angular momentum and red-shifted energy exchanged
magnetically between the hole and the disk. I found that both of these
quantities grow rapidly and that their deposition becomes highly 
concentrated near the inner edge of the disk as the black hole spin 
is increased.

Finally, note that, in the case of an open or partially-open field 
configuration responsible for the Blandford--Znajek process, one 
has to consider magnetic field lines that extend from the event 
horizon out to infinity. Since these field lines are not attached 
to a heavy, infinitely-conducting disk, their angular velocity 
$\Omega_F(\Psi)$ cannot be explicitly prescribed; it becomes just 
as undetermined as the poloidal current~$I(\Psi)$ they carry. 
Fortunately, however, these field lines now have to cross two 
light cylinders, each of which being a singular surface of the 
Grad--Shafranov equation. Therefore, one can impose corresponding 
regularity conditions on both of these two surfaces and use them
in some coordinated, self-consistent manner to determine the two 
free functions $\Omega(\Psi)$ and $I(\Psi)$ simultaneously as a 
part of the overall iterative scheme.

As a final disclaimer, just in case one wonders, the subject of the
present paper has nothing to do with the role of thermal conduction 
in galaxy clusters~\cite{Zakamska-2003} or with the excitation and 
propagation of eccentricity disturbances in planetary systems~\cite
{Zakamska-2004}.

This research was supported by the National Science Foundation 
under Grants Nos.~PHY99-07949 and~PHY-0215581.

~


\bigskip 


\end{document}